# Evolution of the crystal structure and magnetic properties of Sm-doped BiFeO$_3$ ceramics across the phase boundary region

D.V. Karpinsky[a], A. Pakalniškis[b],

G. Niaura[c], D.V. Zhaludkevich[a], A.L. Zhaludkevich[a], S.I. Latushka[a], M. Silibin[d],

M. Serdechnova[e], V.M. Garamus[e], A. Lukowiak[f], W. Stręk[f], M. Kaya[g],

R. Skaudžius[b], A. Kareiva[b]

[a]*Scientific-Practical Materials Research Centre of NAS of Belarus, 220072 Minsk, Belarus*

[b]*Institute of Chemistry, Vilnius University, LT-03225 Vilnius, Lithuania*

[c]*Department of Organic Chemistry, Center for Physical Sciences and Technology, LT-10257, Vilnius, Lithuania*

[d]*National Research University of Electronic Technology "MIET", 124498 Moscow, Russia*

[e]*Helmholtz-Zentrum Geesthacht Centre for Materials and Coastal Research, DE-21502 Geesthacht, Germany*

[f]*Institute of Low Temperature and Structure Research, Polish Academy of Sciences, PL-50422 Wroclaw, Poland*

[g]*Ankara University, Institute of Accelerator Technologies, 06830 Golbasi, Ankara, Turkey*

Corresponding authors:

A. Pakalniškis (pakalniskis.andrius@chgf.vu.lt); D. Karpinsky (dmitry.karpinsky@gmail.com).

**Abstract**

Samarium doped BiFeO$_3$ compounds having nano-size crystallites were prepared by the ethylene glycol assisted sol-gel synthesis method. X-ray diffraction and SEM measurements as well as Raman spectroscopy and FTIR experiments were used to clarify an evolution of the crystal structure on microscopic and local scale levels in the compounds having formula Bi$_{1-x}$Sm$_x$FeO$_3$, where $0 \leq x \leq 1$. Magnetization measurements along with the structural data were used to determine a correlation between magnetic properties and crystal structure of the compounds. The compounds with $0.1 \leq x \leq 0.2$ are characterized by the sequence of the structural transitions driven by the dopant increase, viz. from the polar rhombohedral phase to the non-polar orthorhombic phase via two-phase regions characterized by the presence of the anti-polar orthorhombic phase. The concentration regions ascribed to a coexistence of the rhombohedral and the anti-polar orthorhombic phases as well as the anti-polar orthorhombic and the non-polar orthorhombic phases are observed in the ranges $0.12 \leq x \leq 0.15$ and $0.15 < x < 0.18$ respectively. The compounds having single phase rhombohedral structure are characterized by a release of remnant magnetization as compared to ceramics with similar chemical compositions but synthesized by the solid-state reaction method and having microscopic size crystallites. A correlation between the type of structural distortion, morphology of crystallites and an onset of remnant magnetization is discussed highlighting a difference in the evolution of crystal structure and magnetization observed for the Sm-doped ceramics prepared by modified sol-gel method and conventional solid-state reaction technique.

## 1. Introduction

Over the last decades, ferrites and manganites have attracted great attention of scientific society [1–7], especially the multiferroic $BiFeO_3$ due to the complex interplay between the ferroic orders and the possibility to control the properties using various chemical doping schemes and synthesis methods [8–12]. Chemical doping can be used as an effective tool to neutralize the known drawbacks of the $BiFeO_3$-based materials – high leakage current, nullified remanent magnetization, low or canceled magnetoelectric coupling [13–18]. Chemical doping can also be used to induce metastable structural state observed for the compounds in the vicinity of the structural phase transitions and characterized by an increased sensitivity to the external stimuli as temperature, pressure, electric and magnetic field etc. [19–21].

Chemical substitution of bismuth ions by rare-earth elements (La - Sm) causes the structural transition from the polar rhombohedral phase (described by the space group *R3c, R-phase*) to the non-polar orthorhombic phase (s.g. *Pnma, $O_1$-phase*) via stabilization of $PbZrO_3$-like anti-polar orthorhombic phase (*$O_2$-phase*) [21–23]. The concentration ranges of the structural stability of the rhombohedral and the anti-polar orthorhombic phases strongly depend on the type of rare-earth element and the reduction in the ionic radius of the dopant ions leads to a shrinkage of the mentioned ranges [24,25]. The concentration range attributed to the anti-polar orthorhombic phase reduces down to 1% in the system $Bi_{1-x}Sm_xFeO_3$, while a chemical doping with rare earth elements having ionic radius smaller than that specific for samarium ions leads to a structural transition directly to the non-polar orthorhombic phase [26–28]. Chemical doping with samarium ions attracts particular attention due to very narrow concentration range ascribed to the single phase anti-polar orthorhombic state. It is known that the concentration ranges attributed to the phase coexistence regions can be modified using different synthesis methods and preparation conditions thus allowing a stabilization of relaxor -type ferroelectric behavior and/or diluted magnetic structure [25,30]. In the case of Sm-doping the concentration range scribed to structural stability of the antipolar orthorhombic phase becomes extremely dependent on the preparation conditions and post synthesis treatment of the samples [22,29]

It should be noted that a formation of the metastable structural state in the compounds having frustrated ferroic orders opens up new possibilities for practical applications of the $BiFeO_3$-based multiferroic materials [31,32]. In the present study, we report on the correlation between the type of structural distortion, morphology of crystallites and an onset of remanent magnetization for Sm-doped $BiFeO_3$ compounds prepared by different methods. The obtained results highlight the difference in the concentration driven phase transitions observed for the sol-gel compounds having nano-size crystallites as compared to the ceramic compounds with the same chemical compositions but having microscopic size crystallites due to solid-state reaction method of synthesis.

## 2. Experimental

Analytical grade chemicals of $Bi(NO_3)_3 \cdot 5H_2O$, $Fe(NO_3)_3 \cdot 9H_2O$, $Sm_2O_3$ and ethylene glycol were used as starting materials. Oxide $Sm_2O_3$ was firstly dissolved in 50 ml solution of distilled water containing 2 ml conc. $HNO_3$. The temperature of the solution was raised to 80 °C and then the remaining metal precursors were added, while constantly stirring. After all of the precursors had dissolved, the solution was left to stir for 1 h. Then a required amount of $C_2H_6O_2$ was added to the solution, with a ratio of 2:1 to metal ions. The addition of $C_2H_6O_2$ was followed by a further 1 h of stirring and then finally evaporated at 200 °C (temperature of the magnetic stirrer). Obtained gel was then dried at 150 °C overnight in a drying furnace. Then xerogel was ground in an agate mortar and heated in a furnace at 800 °C for 1.5 h with a heating rate of 1°C/min.

X-ray diffraction (XRD) measurements were performed using Rigaku MiniFlex diffractometer using Cu Kα radiation in the 2Theta range 10º - 70º with a step of 0.02º. Scanning electron microscopy (SEM) images were taken using Hitachi SU-70 SEM. Raman spectra were recorded using inVia Raman (Renishaw, United Kingdom) spectrometer equipped with a thermoelectrically cooled (-70 °C) CCD camera and a microscope. Raman spectra were excited with 532 nm beam from the CW diode pumped solid-state (DPSS) laser (Renishaw, UK). Fourier Transform Infrared (FTIR) absorption spectra were recorded using Shimadzu IRTracer-100 Spectrophotometer in the wavenumber range 350–4500 $cm^{-1}$ with a resolution of 4 $cm^{-1}$ with an attenuated total reflectance (ATR) accessory. Magnetic measurements were performed using Physical Properties Measurement System (Cryogenic Ltd., UK) in magnetic fields up to 14 Tesla in the temperature range 5 – 300 K.

3. **Results and discussion**

*3.1 X-ray diffraction and SEM measurements*

X-ray diffraction data obtained for the compounds $Bi_{1-x}Sm_xFeO_3$ have testified a formation of a continuous series of solid solutions in the whole concentration range ($0 \leq x \leq 1$). The XRD patterns recorded for the different solid solutions presented as the contour maps (Fig.1) show an evolution of the diffraction reflections associated with the changes in the crystal structure occurred in the compounds PDF (COD 96-152-6442). The XRD patterns of the compounds with $x \leq 0.12$ were successfully refined using the single-phase model with the rhombohedral structure (space group *R3c*). An increase in the dopant content leads to the appearance of the reflections specific for the anti-polar orthorhombic phase (s.g. *Pbam*), e.g. $(110)_{O2}$, $(130)|(112)_{O2}$, $(210)_{O2}$ located at 2Θ ~ 27.7°, 29.0° and 33.1° respectively (Fig. 2, insets). It should be noted that the space group *Pbam* (#55) chosen to describe antipolar orthorhombic distortion assumes a doubling of the *c*-parameter as compared to the space group *Pnam* (#6205) used to describe La-doped $BiFeO_3$ compounds [21]. The unit cell of the latter compounds is characterized by a quadrupling of *c*-parameter which is described by the metric $\sqrt{2}a_p * \sqrt{2}a_p * 2\sqrt{3}a_p$, where $a_p$ – fundamental perovskite lattice parameter [25]. The intensity of the reflections attributed to the anti-polar orthorhombic phase gradually increases with the dopant content while the peaks specific for the rhombohedral distortion, e.g. $(113)_R$ and $(006)_R$ (Fig. 2, inset) gradually decrease denoting a reduction of the volume fraction of the rhombohedral phase. The concentration range attributed to the coexistence of the polar rhombohedral and the anti-polar orthorhombic phases is located in the range $0.12 \leq x \leq 0.15$, quite narrow concentration interval points at high chemical homogeneity of the compounds and approves the optimal synthesis conditions used for the sample preparation. Further increase in the dopant content leads to the appearance of the new X-ray diffraction peaks ascribed to the non-polar orthorhombic structure described by the space group *Pnma* (#62) which is typical choice for heavily doped $Bi_{1-x}Re_xFeO_3$ compounds [25, 26].

It should be noted than the reflections attributed to the rhombohedral phase become to be negligible in the compounds with $x > 0.15$ whereas the reflections specific for the anti-polar orthorhombic phase are notable in the XRD patterns of the compounds with the dopant content up to 18 mol%. Thus, the concentration range attributed to the coexistence of the anti-polar and the non-polar orthorhombic phases is located in the range $0.15 < x < 0.18$.

The compounds having the dopant content x > 0.18 were successfully refined assuming the single-phase structural state with the non-polar orthorhombic structure (s.g. *Pnma*). Generally, the sequence of the phase transitions observed for the compounds $Bi_{1-x}Sm_xFeO_3$ is characteristic for the concentration driven phase transitions observed in the compounds $Bi_{1-x}RE_xFeO_3$ (RE = La – Sm) while there are some notable differences discussed below.

Rietveld analysis performed for the XRD patterns of the compounds across the phase boundary region allowed to determine the structural parameters of the different structural phases as well as to itemize a modification in the volume ratio of the phases as a function of the dopant content. In contrast to the compound $Bi_{0.88}Sm_{0.12}FeO_3$ having dominant a rhombohedral phase and small amount (~8%) of the $O_2$-phase the XRD pattern of the compound with x = 0.14 testifies a dominance of the anti-polar orthorhombic phase with minor (~20%) amount of the polar rhombohedral *R*-phase. While the compound with x = 0.16 is already characterized by a formation of the non-polar orthorhombic phase and the structural state of the compound is a coexistence of the dominant anti-polar orthorhombic phase (~77%) with a notable fraction of the non-polar orthorhombic phase (Table #1). It should be noted that with an accuracy of the experimental conditions we could not determine a compound having single phase anti-polar orthorhombic structure as well as we could not reveal any traces of the rhombohedral phase in the diffraction pattern of the compound $Bi_{0.84}Sm_{0.16}FeO_3$. Further increase in the dopant content leads to an increase in the volume fraction of the non-polar orthorhombic state (Table #1) and a formation of the single-phase state in the compound with x = 0.2. It should be noted that monotonous reduction in the unit cell parameters and the lattice volumes of the different structural phases, as well as small (less than 2%) difference in the unit cell volumes ascribed to the different structural phase of two-phase compounds approve an intrinsic character of the phase separation occurred in the chemically homogeneous solid solutions.

It is known that chemical substitution along with the synthesis conditions can cause a modification of crystallites morphology and thus affect the concentration and temperature ranges of the structural stability of different phases [33,34]. In order to trace a modification in the chemical compositions and morphology of the crystallites the SEM measurements were performed. The SEM micrographs of the compounds $Bi_{1-x}Sm_xFeO_3$ are presented in the Fig. 3 Fig. 3 contains the SEM images of the compounds with $0 \leq x \leq 1$ and shows the evolution of the grain size along whole series of the solid solutions. An average grain size estimated for the compounds with $x < 0.4$ is about 300 - 400 nm, however their size distribution range is rather broad and is from 0.1 to 1.4 μm. Increase in the dopant content leads to a reduction in the average grain size and narrowing of the size distribution range.

For the compounds with $0.4 < x < 0.8$ an average grain size reduces from ~ 0.3 μm for $x = 0.4$ down to ~ 0.1 μm for $x = 1$ and the decrease which is accompanied with drastic shortening of the grain size distribution range. It should be noted that a decrease in the average size of the grain with the dopant content is associated with reduced chemical reactivity of Sm ions as compared to Bi ions thus leading to a reduction in the mass transport becomes slower [35]. A smaller amount of the oxygen vacations formed due to evaporation of bismuth ions [36] also leads to a reduction in the grain size in the heavily doped compounds. Analysis of the SEM images obtained for the compounds with the chemical compositions across the structural transitions from the rhombohedral to the orthorhombic structure $0.1 < x \leq 0.2$ testifies a modification in the grain form while the size remains nearly stable. The compounds having dominant rhombohedral structure are characterized by rounded shape grains, increase in the dopant content leads to a formation of the grains with the distinct rectangular-like shape which corresponds to the geometry of the unit cells attributed to the different structures [37]. The compounds with $x \geq 0.6$ are characterized by nanoscale size grains with semispherical shape which becomes to be energetically favorable for nanometer particles [38,39].

*3.2 FTIR and Raman measurements, phase diagrams*

The evolution of the structural phases was also analyzed using the data obtained by the local scale measurements performed by Raman and FTIR spectroscopy which gave complementary structural data to the results of the X-ray diffraction measurements. The FTIR spectra presented in Fig. 4 allowed to trace a modification of the vibration modes attributed to the different structural states across the phase boundary region. Strong peaks located at ~ 420 and 540 $cm^{-1}$ are ascribed respectively to stretching and bending vibrations occurred in $FeO_6$ octahedra in the rhombohedral structure [40–42]. An increase in the dopant content up to 14% leads to a formation of the band around 680 $cm^{-1}$ which is associated with a modification of dipole moment specific for the anti-polar orthorhombic structure. The intensity of the mentioned band reduces with Sm content and becomes to be negligible in the compound with x = 0.18 thus confirming the X-ray diffraction data which shows a stabilization of the anti-polar orthorhombic phase in the concentration range $0.12 \leq x \leq 0.18$. The FTIR spectra of the compounds with x > 0.18 are characterized by the active band at ~ 510 cm$^{-1}$ which is specific for bending vibrations O – Fe – O in the non-polar orthorhombic structure [43].

Diffraction studies afford important knowledge about the average crystal structure of studied compounds, while Raman spectroscopy provides molecular level information on short range structure or local symmetry which is difficult to acquire by other structure-sensitive techniques [44]. In addition, Raman spectroscopy is very sensitive to the presence of defects and disorders [45–47]. Fig. 5 compares 532-nm excited Raman spectra of polycrystalline BiFeO$_3$ and SmFeO$_3$ samples. The strong high frequency features located at 1261 and 1288 cm$^{-1}$ for BiFeO$_3$ and SmFeO$_3$ compounds, respectively, belong to the two-phonon vibrational mode, which can be described as an overtone of oxygen stretching vibration [48]. The well-defined characteristic bands of rhombohedral BiFeO$_3$ structure are visible at 78 cm$^{-1}$ (symmetry $E$), 141 cm$^{-1}$ ($A_1$), 173 cm$^{-1}$ ($A_1$), and 221 cm$^{-1}$ ($A_1$) [45,49]. The low intensity peaks at 469 and 527 cm$^{-1}$ belong to $A_1$ and $E$ symmetry vibrational modes, respectively [49]. The origin of low intensity band near 604 cm$^{-1}$ is not completely clear; it might be associated with second-order vibrational mode or $E$ symmetry longitudinal optical (LO) phonon [50–52]. First principles calculations predict vibrational modes associated with a motion of mainly Bi$^{3+}$ ion at frequencies lower than 167 cm$^{-1}$, while the bands visible at frequencies higher than 262 cm$^{-1}$ are associated with motion of oxygen atoms [50].

A different spectral pattern was observed for SmFeO$_3$ compound (Fig. 5). The main peak positions correspond well to the orthorhombic *Pnma* structure of SmFeO$_3$ [52]. Theoretical analysis predicts 24 Raman active vibrational modes for such structure ($7A_g + 5B_{1g} + 7B_{2g} + 5B_{3g}$) [52]. The main atomic motions associated with particular vibrational modes are described in-details by Weber et al. [52]. Three low-frequency bands located at 109 cm$^{-1}$ [$A_g(1)$], 143 cm$^{-1}$ [$A_g(2)$], and 157 cm$^{-1}$ [$B_{2g}(2)$] are related with motion of Sm$^{3+}$ ion. The middle intensity bands near 233 cm$^{-1}$ [$A_g(3)$] and 376 cm$^{-1}$ [$A_g(5)$] are related mainly with in-phase rotation of FeO$_6$ group. The well-defined intense band at 315 cm$^{-1}$ [$A_g(4)$] is associated mainly with a motion of O(1) atom. The middle intensity band located at 422 cm$^{-1}$ [$A_g(6)$] is related mainly with in-phase Fe-O(2) stretching vibration. Finally, the well-defined band at 465 cm$^{-1}$ [$B_{2g}(5)$] belongs to O(1)-Fe-O(2) scissoring-like bending vibration. The broad feature near 635 cm$^{-1}$ probably is associated with second-order vibrational mode [52].

Fig. 6 shows composition-induced changes in the Raman spectra of Bi$_{1-x}$Sm$_x$FeO$_3$ compounds in the frequency region of 70−750 cm$^{-1}$. Even a presence of a small amount of Sm$^{3+}$ ions (x = 0.08) results in an observable shift of the vibrational bands; three characteristic BiFeO$_3$ modes located at 141, 173, and 221 cm$^{-1}$ shifted to higher wavenumbers by 4, 3, and 11 cm$^{-1}$, respectively. The progressive upshift frequency of these bands is visible also after further increasing amount of Sm$^{3+}$ ions (x = 0.12). Changes in peak positions point to a perturbation in the lattice structure; however, the general spectral pattern remains similar, indicating that the rhombohedral structure is preserved.

More general structural perturbations are visible at composition x = 0.14. Further replacement of $Bi^{3+}$ ions by $Sm^{3+}$ results in drastic changes in the lattice structure (x = 0.20); characteristic low-frequency bands of rhombohedral structure disappear and broad feature near 303 cm$^{-1}$ becomes visible. Such a new structure dominates in the composition range from x = 0.20 to 0.60. At x = 0.80 spectral features characteristic to $SmFeO_3$ start to appear.

Based on the structural data obtained by X-ray diffraction and the results of the local scale measurements (Raman, FTIR) as well as the previously published data [22] the concentration dependent phase diagrams were constructed for the sol-gel compounds and the solid-state ceramics (Fig. 7). The phase diagrams represent the single phase and the two-phase regions. The structural transition from the rhombohedral to the anti-polar orthorhombic phase occurs in the sol-gel compounds in a narrow concentration range (~3%) which is consistent with a small (~0.4%) difference in the reduced unit cell volumes as confirmed by the XRD data. The transition to the non-polar orthorhombic state lasts in the concentration range of about 5% and the average difference in the reduced unit cell volumes is more pronounced (about 1.8%).

In the solid-state ceramics similar sequence of the phase transitions occurs within nearly the same concentration range (0.11 < x < 0.18) and related changes are observed for the unit cell parameters [22]. The main difference in the evolution of the structural phases occurred in the compounds prepared by different synthesis methods is a stabilization of the single-phase structural state with the anti-polar orthorhombic structure in the solid-state compound with x ~ 0.14. The anti-polar orthorhombic structure is considered to be the universal intermediate phase in the concentration- and temperature- driven phase transition in the $BiFeO_3$-based compounds [25]. The absence of the single phase anti-polar orthorhombic structure is most probably caused by the reduced crystallite size and the grain morphology formed in the sol-gel compounds as compared to microscopic size crystallites specific for solid-state ceramics [25,53]. The crystallites size and the grain morphology affect not only the sequence of the structural phase transitions but also the magnetic properties of the compounds as discussed in the next section.

*3.3 Magnetization measurements*

Magnetization measurements performed for the compounds $Bi_{1-x}Sm_xFeO_3$ have allowed to trace the evolution of magnetic properties as a function of the dopant content as well as to clarify a correlation between different structural states and a release of remanent magnetization. The presented magnetization data are related to the phase boundary region ($0.08 \leq x \leq 0.2$) characterized by a set of structural transitions. The isothermal magnetization dependencies (Fig. 8) recorded for the compound with 8 mol.% of samarium content testify a partial disruption of the modulated magnetic structure specific for the initial compound $BiFeO_3$ [54] which is evidenced by a kink at the M(H) dependence observed at the magnetic field of ~5 T [55]. An increase in the dopant content leads to a complete disruption of the modulated structure and weak ferromagnetic state stabilizes. Small remanent magnetization observed in the compounds with $x > 0.08$ is associated with spin canting caused by Dzyaloshinskii-Moriya interaction as well as structural defects and chemical inhomogeneities which also increase non-zero magnetization [25,56].

Compound $Bi_{0.88}Sm_{0.12}FeO_3$ is characterized by a remanent magnetization of about 0.14 emu/g which is nearly two times larger than the value of remanent magnetization measured for the same compound prepared by solid-state reaction method and also having nearly single phase rhombohedral structure [15,19]. Sol-gel compound with x = 0.14 is characterized by a dominant anti-polar structure and spontaneous magnetization of about 0.26 emu/g which is about the value specific for the solid-state compound. It should be noted that the coercivity of the sol-gel compound is two times larger than that in the solid-state compound which is caused by a mixed magnetic state formed in the sol-gel sample. Further chemical doping leads to a stabilization of magnetic structure associated with weak ferromagnetism and spontaneous magnetization remains nearly stable ($M_R \sim 0.28$ emu/g) for the compounds having either single phase orthorhombic structure or the two-phase structural state. The coercivity of the compounds drastically increases with the dopant content (up to ~3.5 T for x = 0.2) which is most probably caused by a rapid decrease in the grain size and thus a modification of magnetic structure from multi domain to single domain state [57]. It should be noted that evolution of the magnetic structure is quite monotonous in spite of the structural transitions from the polar rhombohedral phase to the non-polar orthorhombic phase via the anti-polar orthorhombic one. The main difference in the magnetization of the sol-gel compounds as compared to the magnetization of the solid-state ceramics is related to a rapid increase in remanent magnetization with the dopant content in the compounds having single phase rhombohedral structure while in the solid-state ceramics a release of remanent magnetization was associated with an onset of the orthorhombic phase [15,19].

## 4. Conclusions

Based on the obtained results we conclude that in the compounds $Bi_{1-x}Sm_xFeO_3$ ($0 \leq x \leq 1$) prepared by a modified sol-gel method the single-phase rhombohedral structure is stable up to $x < 0.12$. An increase in the concentration of samarium leads to the appearance and a rapid increase in the intermediate anti-polar orthorhombic phase and the compounds with $0.12 \leq x \leq 0.15$ are characterized by a coexistence of the rhombohedral and the anti-polar orthorhombic phases. The concentration range $0.15 < x \leq 0.18$ is ascribed to the two-phase compounds characterized by the coexistence of the anti-polar and the non-polar orthorhombic phases. In contrast to the structural results obtained for the solid-state ceramics the single-phase state with anti-polar orthorhombic structure is not stabilized in the compounds under study. The compounds having single phase rhombohedral structure are characterized by a notable release of remanent magnetization in contrast to the solid-state compounds having the same chemical compositions. The obtained results testify the absence of a distinct correlation between the type of structural distortion and a release of remanent magnetization as it was considered for the compounds $Bi_{1-x}RE_xFeO_3$ prepared by solid-state method. The difference in the correlation between the structure and magnetic properties of the compounds prepared by different methods is mainly caused by significant contribution to non-zero magnetization from the surface layer of nano-size grains formed in the sol-gel compounds.

## Acknowledgements

This work was supported by the European Union's Horizon 2020 research and innovation programme under the Marie Skłodowska-Curie grant agreement No. 778070. G.N. gratefully acknowledges the Center of Spectroscopic Characterization of Materials and Electronic/Molecular Processes (SPECTROVERSUM Infrastructure) for use of Raman spectrometer. M.K. acknowledges the TARLA project founded by the Strategy and Budget Department of Turkey (project code: 2006K12-827). M.S. and D.K. acknowledge RFBR (projects # 20-58-00030 and 18-38-20020 mol_a_ved) and BRFFR (project # F20R-123).

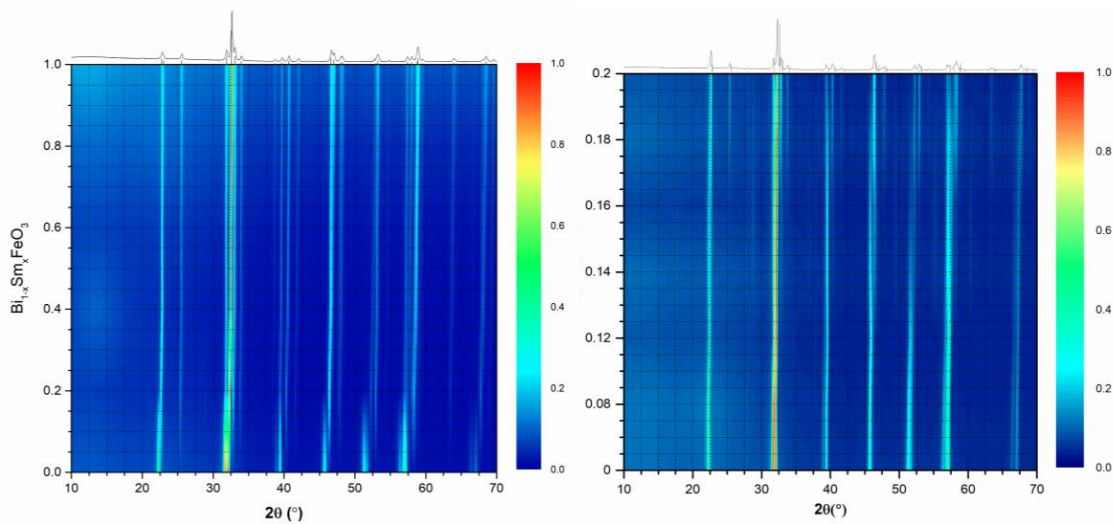

Fig. 1. Room temperature XRD patterns of the compounds $Bi_{1-x}Sm_xFeO_3$; left image denotes the compounds with $0 \leq x \leq 1$ – left image; the results obtained for the compounds with $0 \leq x \leq 0.2$ - right image. Blue color indicates the lowest intensity (background), red color represents the most intensive points (the peaks).

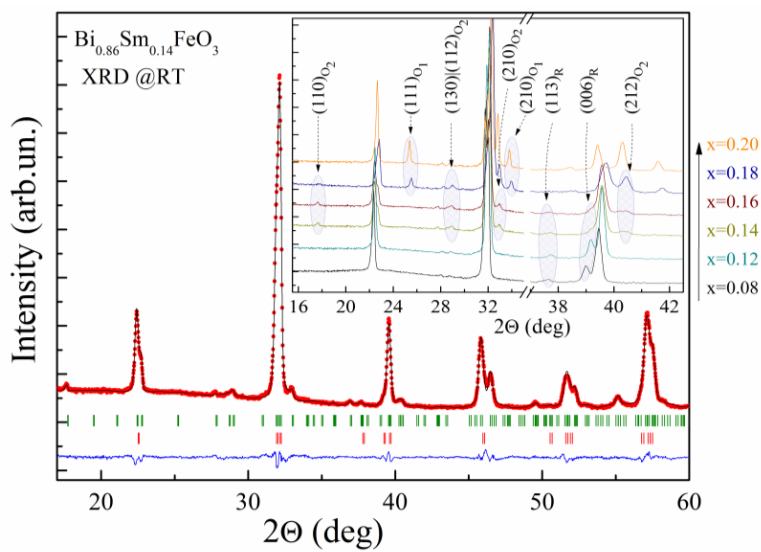

Fig. 2. The XRD pattern of $Bi_{0.86}Sm_{0.14}FeO_3$ recorded at room temperature (red dots are the experimental data; black lines are calculated ones). Bragg reflections were indicated by vertical ticks. The inset shows the concentration driven evolution of the reflections ascribed either to the *R*- or $O_2$- phases.

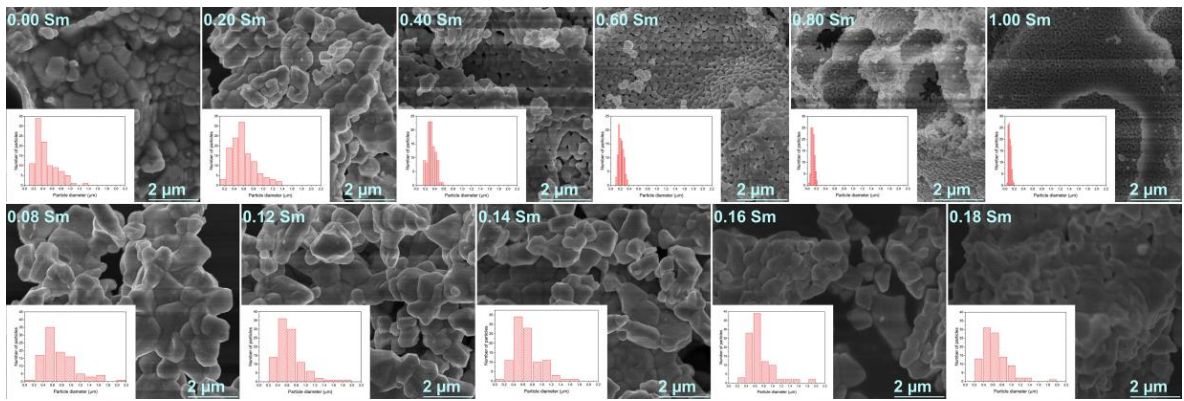

Fig. 3. SEM images of $Bi_{1-x}Sm_xFeO_3$ solid solutions with $0 < x < 1$, insets represent the particle size and their size distribution.

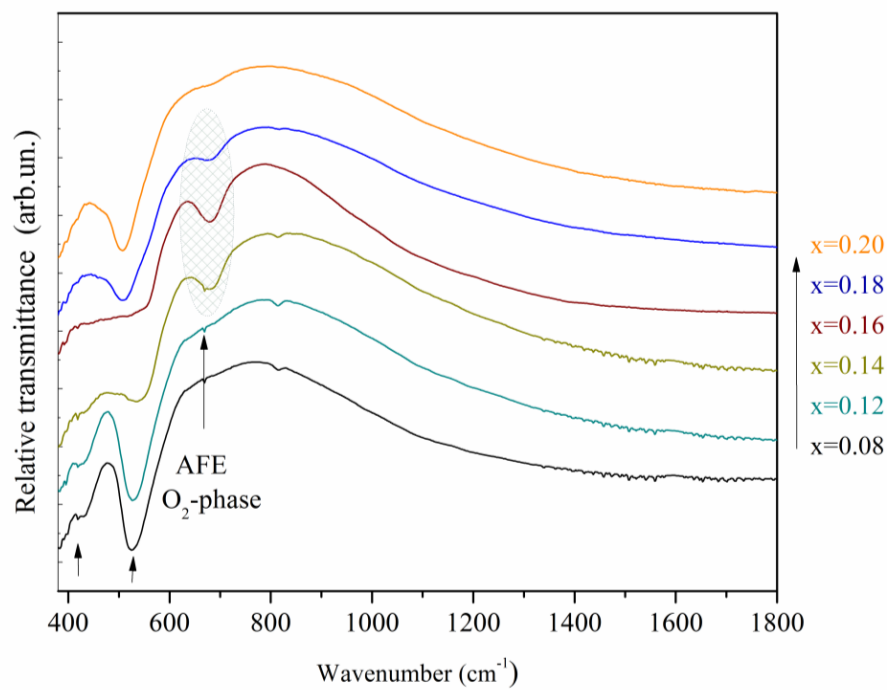

Fig. 4. FTIR spectra of the compounds $Bi_{1-x}Sm_xFeO_3$ with $0.08 \leq x \leq 0.2$.

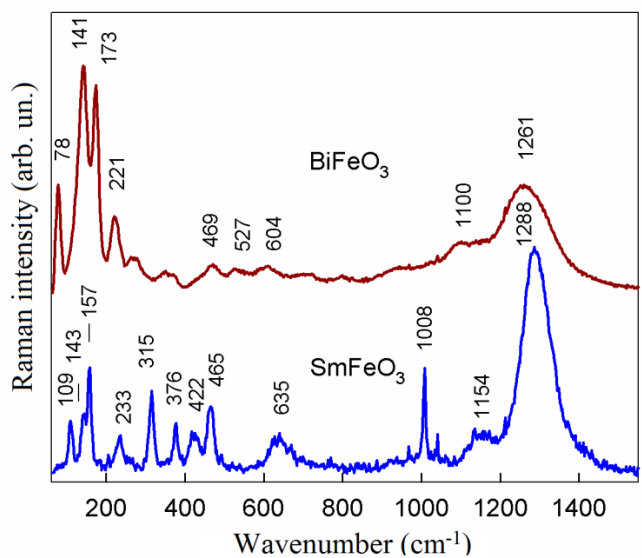

Fig. 5. Raman spectra of polycrystalline $BiFeO_3$ and $SmFeO_3$. The excitation wavelength is 532 nm (0.06 mW).

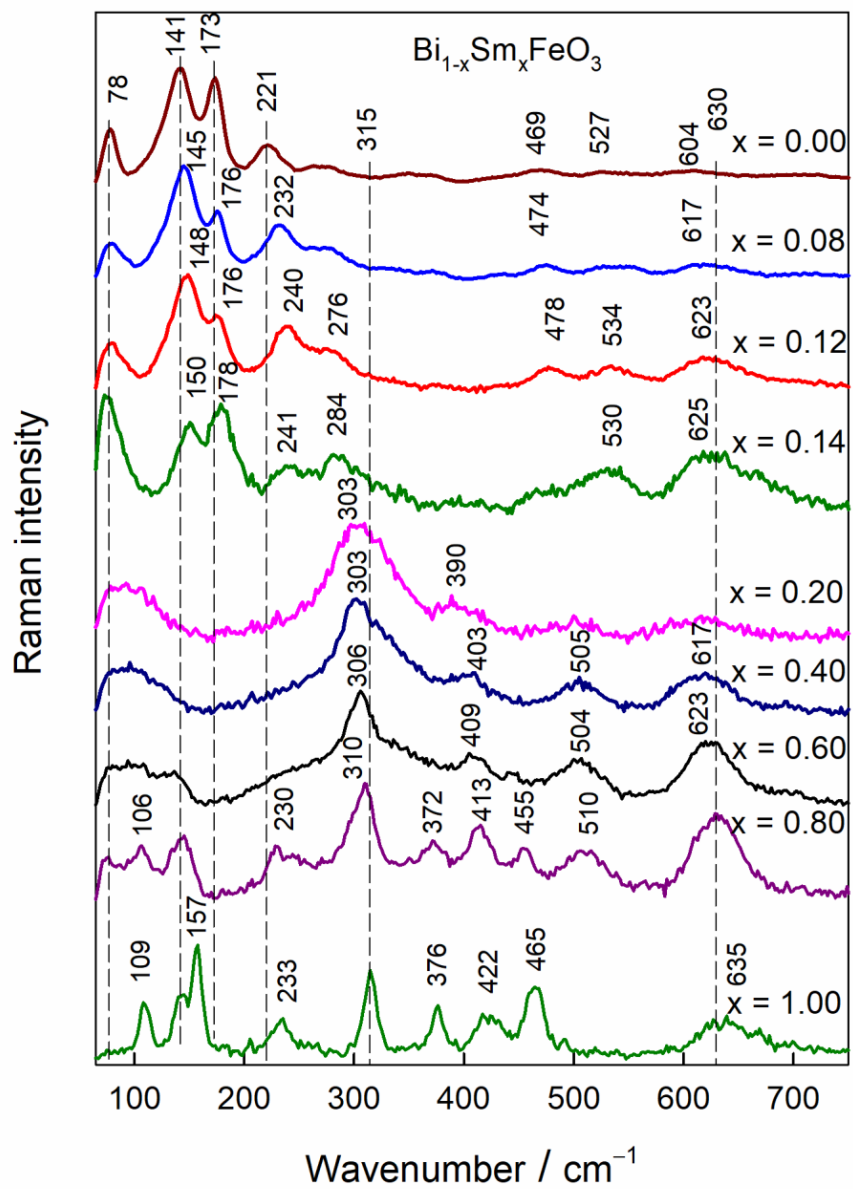

Fig. 6. Composition dependent Raman spectra of polycrystalline $Bi_{1-x}Sm_xFeO_3$ compounds. Intensities are normalized to the intensity of the most intense band and spectra are shifted vertically for clarity. The excitation wavelength is 532 nm (0.06 mW).

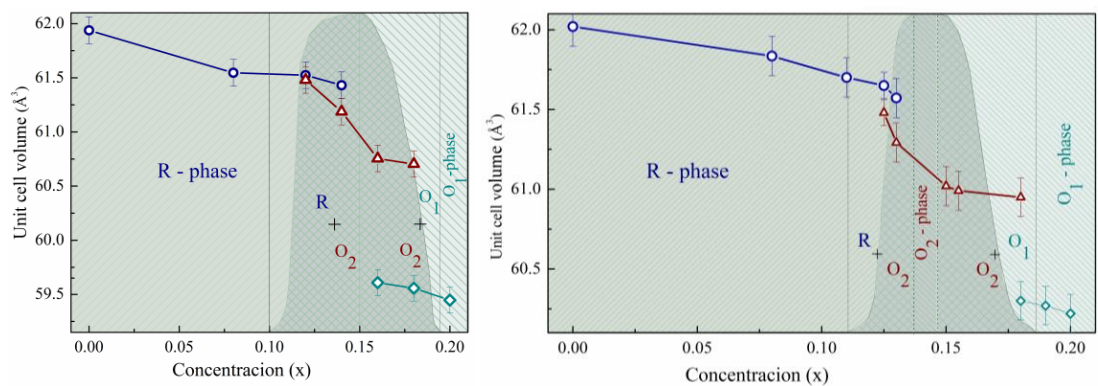

Fig. 7. Phase diagrams of the compounds $Bi_{1-x}Sm_xFeO_3$ prepared by sol-gel (left) and solid-state (right) reaction technique.

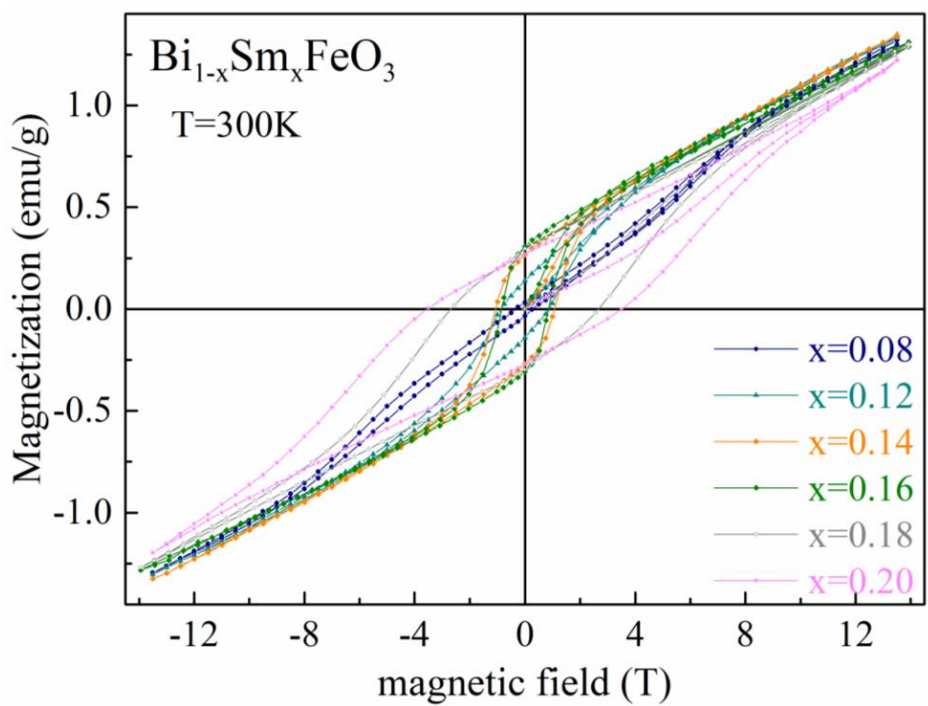

Fig. 8. Magnetic field dependencies obtained for the compounds $Bi_{1-x}Sm_xFeO_3$ at room temperature.

Table 1. The reduced unit cell parameters and the structural phases ratios calculated as well coercivity with saturation magnetization for the compounds $Bi_{1-x}Sm_xFeO_3$ with chemical compositions across the phase boundary region ($0.08 \leq x \leq 0.2$).

| Sample | Phase | $a$, Å | $b$, Å | $c$, Å | Volume Å$^3$ (per primitive cell) | Coercivity, T | Saturation magnetization, emu/g |
|---|---|---|---|---|---|---|---|
| $Bi_{0.92}Sm_{0.08}FeO_3$ | R3c | 3.932(1) | 3.932(1) | 3.980(7) | 61.54(6) | 0.2 | 1.29 |
| $Bi_{0.88}Sm_{0.12}FeO_3$ | R3c (92%) | 3.933(2) | 3.933(2) | 3.976(4) | 61.52(5) | 0.9 | 1.33 |
| | Pbam (8%) | 3.961(3) | 3.969(4) | 3.909(7) | 61.48(5) | | |
| $Bi_{0.86}Sm_{0.14}FeO_3$ | R3c (20%) | 3.932(7) | 3.932(7) | 3.973(6) | 61.43(5) | 1.1 | 1.34 |
| | Pbam (80%) | 3.953(5) | 3.963(9) | 3.904(6) | 61.18(3) | | |
| $Bi_{0.84}Sm_{0.16}FeO_3$ | Pbam (77%) | 3.943(4) | 3.956(7) | 3.893(4) | 60.75(1) | 0.9 | 1.31 |
| | Pnma (23%) | 3.960(1) | 3.910(2) | 3.840(1) | 59.60(7) | | |
| $Bi_{0.82}Sm_{0.18}FeO_3$ | Pbam (32%) | 3.937(5) | 3.955(9) | 3.898(6) | 60.70(9) | 2.7 | 1.28 |
| | Pnma (68%) | 3.962(3) | 3.907(6) | 3.846(3) | 59.55(7) | | |
| $Bi_{0.80}Sm_{0.20}FeO_3$ | Pnma | 3.967(9) | 3.900(2) | 3.841(8) | 59.44(4) | 3.5 | 1.22 |